\documentclass[letterpaper,twocolumn,prd]{revtex4}

\usepackage{epsfig}
\def\cadez{\v{C}ade\v{z}}

\keywords{regge calculus; numerical relativity}
\preprint{LAUR-01**}

\newcommand{\doublefigure}[2]{
  \begin{center}
    \begin{tabular}{cc}
      \epsfig{file=#1, width=0.22\textwidth} &
      \epsfig{file=#2, width=0.22\textwidth} \\
      \mbox{\sl(a)} & \mbox{\sl(b)} 
    \end{tabular}
  \end{center}
  }

\newcommand{\quadfigure}[4]{
  \begin{center}
    \begin{tabular}{cc}
      \epsfig{file=#1, width=0.15\textwidth, angle=270} &
      \epsfig{file=#2, width=0.15\textwidth, angle=270} \\
      \mbox{\sl(a)} & \mbox{\sl(b)} \\
      \epsfig{file=#3, width=0.15\textwidth, angle=270} &
      \epsfig{file=#4, width=0.15\textwidth, angle=270} \\
      \mbox{\sl(c)} & \mbox{\sl(d)} 
    \end{tabular}
  \end{center}
  }

\begin{document}

\title{Regge calculus: a unique tool for numerical relativity}
\author{Adrian P. Gentle}
\affiliation{Theoretical Division (T-6, MS B288), 
  Los Alamos National Laboratory, Los Alamos, NM 87545, USA\footnote{Permanent Address:  Department of Mathematics, University of Southern Indiana, Evansville, IN 47712. E-mail:  apgentle@usi.edu}}

\begin{abstract}
  The application of Regge calculus, a lattice formulation of general
  relativity, is reviewed in the context of numerical relativity.
  Particular emphasis is placed on problems of current computational
  interest, and the strengths and weaknesses of the lattice approach
  are highlighted.  Several new and illustrative applications are
  presented, including initial data for the head on collision of two
  black holes, and the time evolution of vacuum axisymmetric Brill
  waves. 
\end{abstract}

\maketitle

\section{Numerical Relativity}

The complexity of the Einstein equations, combined with the sparsity
of relevant analytic solutions, necessitates a range of other tools
with which to explore complex physical scenarios.  These include
series expansions and perturbation techniques, together with numerical
solutions of the fully non-linear equations.  Unfortunately, the
numerical solution of Einstein's equations has proved to be an
exceedingly difficult problem.  Over the last three decades numerous
schemes have been developed, or adapted, to tackle the vast range of
problems which fall within the purview of numerical relativity.

The classic approach, involving a three-plus-one dimensional split of
space and time, was first expounded by Arnowitt, Deser and Misner
(ADM) \cite{mtw}.  This approach is natural in the context of our
Newtonian intuition, and is also directly applicable to computer
simulations. Whilst the traditional ADM approach has dominated
numerical relativity, questions regarding its long-term stability have
lead to the development of many other formulations of the Einstein
equations in three-plus-one dimensions.

Techniques of current interest include those developed by Sasaki and
Nakamura, and later by Baumgarte and Shapiro, known generically as
Conformal ADM (CADM) formulations.  Incorporating insights from York's
analysis of the initial value problem, these algorithms have shown
superior stability properties compared with ADM in some applications.
Other recent formulations are based on symmetric-hyperbolic forms of
the Einstein equations, where it is hoped that the mathematical proofs
of stability and well-posedness confer numerical advantages compared
with techniques whose mathematical structures are more uncertain.  The
recent review by Lehner \cite{lehner01} discusses many of these
issues.

Despite the enormous effort invested in these techniques (and many
others, including characteristic formulations), most modern numerical
relativity codes continue to suffer from problems with long-term
stability and lack of accuracy.  Lack of resolution is only a partial
solution; new insights or techniques seem to be required to overcome
many of these problems.

\section{Regge's formulation of general relativity on a lattice}

Regge calculus \cite{regge61} is a formulation of general relativity
on a piecewise flat simplicial complex, rather than a differentiable
manifold. In general this complex is built from four-simplices, the
four dimensional generalisation of triangles and tetrahedra.  The
interior of each three and four dimensional lattice element is
intrinsically flat, with curvature concentrated on the two-dimensional
faces (triangles).  In $n$-dimensions, curvature resides on lattice
elements of co-dimension two.

Regge calculus would appear to be ideally suited to numerical
simulations; it is an inherently discrete formulation of general
relativity, and complex topologies are easily incorporated.  In
addition to its facility in classical numerical relativity, Regge
calculus has also seen application in the sum-over-histories
formulation of quantum gravity.  In this review we concentrate on the
application of Regge calculus to classical numerical relativity;
details on quantum applications can be found elsewhere
\cite{williams97}.  The remainder of this section is devoted to the
description of classical gravity on a simplicial lattice.

\begin{figure}[t]
  \begin{center}
    \begin{tabular}{cc}
      \epsfig{file=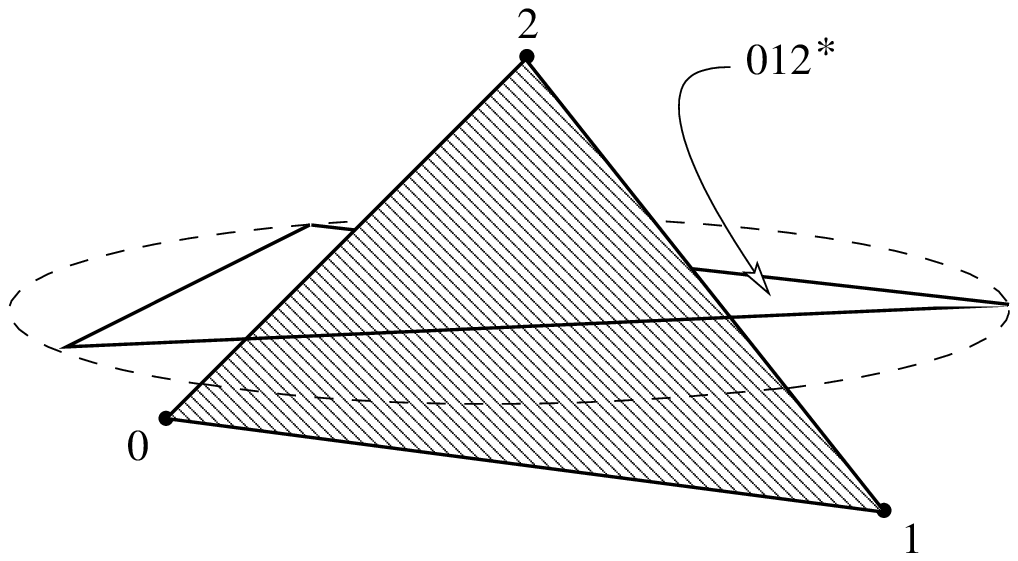, width=0.22\textwidth} &
      \epsfig{file=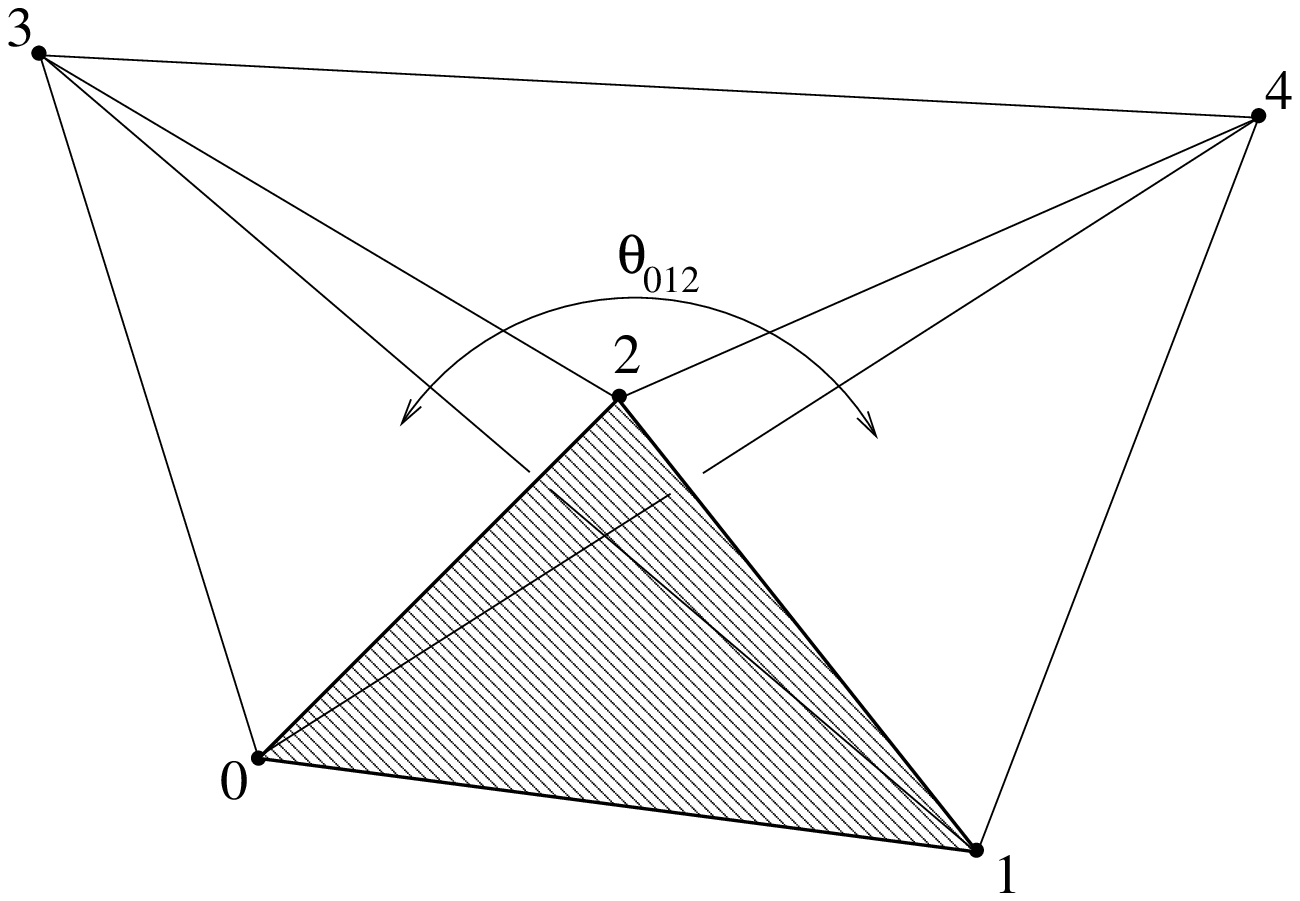, width=0.22\textwidth} \\
      \mbox{{\sl (a)}} & \mbox{{ \sl(b)}}\\
    \end{tabular}
  \end{center}
  \caption{{\sl (a)} A triangle $012$ embedded in a simplicial
    complex (not shown).  To examine the curvature about $012$ one
    transports a vector through a closed loop in the plane orthogonal
    to $012$. The simplicial complex itself naturally defines this
    loop -- the dual triangle, denoted $012^*$.  {\sl (b)} An
    illustration of the hyper-dihedral angle $\theta_{012}$, measured
    between the tetrahedra $0123$ and $0124$ which hinge on $012$
    within the four-simplex $01234$. }
  \label{fig:simplex} 
\end{figure}

Given a lattice spacetime, which consists of a list of vertices, the
corresponding connectivity matrix and the lengths of all edges, the
Gaussian curvature $\kappa$ can be locally defined by the parallel
transportation of a test vector about a closed loop.  Calculating the
angle through which the vector rotates, we obtain
\begin{equation}
  \kappa =  \frac{\mbox{angle rotated}}{\mbox{area of loop}} =
  \frac{\epsilon}{A^*}.
\end{equation}
The first equality is the definition of Gaussian curvature; the second
provides the equivalent expression on the lattice.  The natural loop
to choose, as indicated in figure \ref{fig:simplex}(a), is defined by
the area dual to the triangle on which the curvature resides. This is
a portion of the dual-lattice, and may in principle be calculated for
a given triangulation.  The test vector is rotated through an angle
$\epsilon$ in the plane orthogonal to the triangle (that is, in the
plane of the dual area $A^*$).  We refer to $\epsilon$ as the
``deficit angle'', the lattice representation of the curvature
concentrated on the triangle.  The deficit angle is computed, given
the edge lengths of the lattice, using
\begin{equation}
  \epsilon = 2\pi - \sum_k \theta_{k}
\end{equation}
where the summation is over all four-simplices $k$ which contain the
triangle, and $\theta_k$ is the hyper-dihedral angle between the two
tetrahedral faces which hinge on the triangle within $k$.  Figure
\ref{fig:simplex}(b) shows this structure for the simplex $01234$
which hinges on the triangle $012$.  The form given above applies for
spacetimes with Euclidean signature; similar expressions apply for a
spacetime with signature $-+++$.

Regge derived the simplicial equivalent of the Hilbert action
\cite{regge61} for a lattice spacetime,
\begin{equation}
  \int_{\cal M} \sqrt{-g}\, R\, d^4x \quad \rightarrow \quad 2 \sum
  _{i} \, \epsilon_i A_i
\end{equation}
where the summation is over all triangles $i$, $A_i$ is the area and
$\epsilon_i$ is the deficit angle of the $i$th triangle.  This
simplicial expression can be intuitively understood by noting that on
a simplicial complex the Hilbert integrand has compact support on
triangles, each triangle has the volume element $dV = \sqrt{-g} \, d^4
x \approx A A^*$ and the scalar curvature is $R\approx \kappa =
\epsilon / A^*$.  A more thorough proof in this spirit has been
provided by Miller \cite{miller97}.

Lattice edges are the discrete representation of the metric, and
thus any geometric quantity may be calculated once all edges are
known.  In direct analogy to the continuum, the equations of motion
are obtained by a variational principle; the action is varied with
respect to the independent variables.  This yields the vacuum ``Regge
equations''
\begin{equation}
  0 = \frac{\delta I_R}{\delta l_j} = 2 \sum_i \epsilon_i \frac{\delta
  A_i}{\delta l_j}
\end{equation}
where $I_R$ is the lattice form of the Hilbert action.  These
equations of motion are implied by the Regge identity
\cite{regge61},
\begin{equation}
  \sum_i \frac{\delta \epsilon_i}{\delta l_j} A_i = 0 
\end{equation}
which is the lattice equivalent of the Palatini identity \cite{mtw}
\begin{equation}
  \int_{\cal M}
  g^{\alpha \beta} \delta R_{\alpha \beta} \sqrt{-g} \, d^4 x = 0.
\end{equation}

From the derivation of the lattice equations it is clear that there is
a single equation for each edge in the lattice.  In general we expect
at least seven edges per vertex in a three-dimensional simplicial
lattice (triangulating $R^3$; other topologies may change this average
slightly) compared with the six independent components of the
three-metric per point in the continuum.  This can be understood by
comparison with a tetrad formulation, where one introduces an
extraneous variable and a corresponding consistency condition.  We
expect that the lattice equations themselves provide the additional
conditions.

From this basic structure it is possible, in principle, to construct
the simplicial counterparts of any geometric object of interest:
$K_{\mu\nu}$, $R_{\alpha\beta\mu\nu}$ and so forth.  However, careful
averaging is required if these simplicial definitions are to converge
smoothly (and pointwise) to their continuum counterparts. 

Finally, we note that the theory of Regge outlined above is one among
many conceivable approaches to ``lattice gravity''.  A related method
has recently been developed by Brewin \cite{brewin98}, where the
metric is locally constructed in Riemann normal coordinates from an
underlying lattice structure.  Brewin is then able to use any of the
standard formulations (ADM, CADM, etc) to evolve the lattice.  These
approaches are related in general structure to the finite element and
finite volume methods in wide use in computational fluid dynamics and
engineering.  It is not yet entirely clear how lattice approaches to
general relativity relate to these well-developed numerical
techniques, but such an understanding would provide a useful bridge
between the lattice theory and continuum methods.

\section{(3+1)-dimensional lattice gravity}

The $(3+1)$-dimensional formulation of Regge calculus is reasonably
well understood, although to date there have been few applications.
Significant new studies will be required to obtain a full
understanding of such issues as convergence and long-term stability.

The initial value problem is a vital precursor to the generic
evolution problem.  A general technique for solving the initial value
problem in the thin-sandwich, $(3+1)$-dimensional formulation of Regge
calculus has been described by Gentle and Miller \cite{gentle98}.
Their approach is based on an identification of the geometric degrees
of freedom and conformal structures employed in the York initial value
formalism \cite{york79}.  By associating lattice elements with the
continuum conformal structure, Gentle and Miller describe one possible
(although far from unique) approach to the construction of two-slice
simplicial initial data. They successfully benchmark their approach on
the Kasner cosmology \cite{gentle98}.

Once initial data has been constructed a consistent evolution scheme
is required.  The Regge equations, a set of coupled non-linear
algebraic equations, can in principle be evolved by iteratively
solving the entire coupled system at each timestep.  However, it was
recently realised that the lattice can be structured in such a way as
to allow the implicit, parallel, decoupled evolution of sets of vertices
\cite{committee}.  This algorithm is known as the ``Sorkin evolution
scheme''.

A general $(3+1)$-dimensional lattice may be constructed from a given
three-dimensional lattice by the ``evolution'' (not true
geometrodynamic evolution) of individual vertices.  Each vertex is
``dragged forward'', off the initial hypersurface, and in the process
addition edges are created to join the ``evolved'' vertex to its
counterpart on the initial hypersurface.  This process is repeated
until all vertices have been ``evolved''.  In this way the
connectivity of the initial hypersurface is replicated, whilst the
lattice structure between the two surfaces is constructed naturally as
part of the algorithm.  A lattice of this type, referred to as a
``Sorkin triangulation'' \cite{committee}, allows the parallel
evolution of individual vertices described above.  

General relativity is fundamentally coordinate invariant, although one
must choose a convenient frame in which to perform numerical
calculations.  In the standard $(3+1)$-dimensional ADM formulation,
this is encoded in the freedom to lay down coordinates on the initial
Cauchy surface, together with the freedom to choose how these
coordinates are propagated to each future time slice (lapse and shift
freedom). In a lattice formulation these correspond to the freedom to
distribute vertices on the initial hypersurface and the freedom to
propagate those vertices, respectively.  The latter freedom requires
the provision of four conditions per vertex during each evolution step
to determine the propagation of the vertices; the simplicial lapse and
shift freedom.

Regge calculus has been successfully applied in $(3+1)$-dimensions in
the context of the anisotropic, homogeneous, $T^3$ Kasner cosmology
\cite{gentle98,brewin01}.  These initial studies provide confidence in
the technique, which is able to recover the homogeneity and anisotropy
of the analytic solution, and also demonstrate both stability and
second-order convergence to the continuum solution.

\section{Test-bed Applications}

In the development of any numerical technique one considers a variety
of test problems which provide insight into the strengths and
weaknesses of the approach.  These should be of increasing complexity,
and preferably, cover a wide range of physical scenarios which model,
in a simplified manner, the actual physical problems which motivate
the code development.  In this way one develops confidence in the
solutions obtained when the code is applied to real physical problems.
Such a programme has been under way in the case of Regge calculus since
its inception.  Successful early applications included highly
symmetric test problems with only a few degrees of freedom, together
with the first numerical construction of axisymmetric binary black
hole initial data.  The review by Williams and Tuckey contains a full
discussion of these early studies \cite{williams92}.

More recently Regge calculus has been applied to increasingly complex
problems of both physical and computational import. It is these
studies on which we concentrate in this review.  Successful
applications to the Brill wave \cite{brill59,gentle99a} and black hole
plus Brill wave \cite{bernstein93,gentle99b} initial data sets have
validated computational lattice gravity, whilst also providing
independent confirmation of the physical predictions of previous
finite-difference studies.  In this section we discuss these and other
test-bed applications of the lattice approach, allowing us to
investigate its accuracy and flexibility.

\subsection{Spherical symmetry}

The Schwarzschild solution is a classic test-bed for numerical
relativity.  The combination of non-trivial topology with an event
horizon covering the central singularity has proved a challenging
problem even for modern codes.  In this section we
briefly describe an initial evolution of the static Schwarzschild
solution using lattice gravity.  

The imposition of symmetry conditions in a lattice simulation is
challenging.  While providing enormous flexibility to model complex
topologies, a simplicial lattice (built, in three dimensions, from
tetrahedra) requires careful construction if it is to respect the
spherical symmetry of the single black hole spacetime.  The approach
usually taken involves the construction of a full four-dimensional
geometry constructed from simplices, after which the lattice is
collapsed along the symmetry axes to obtain spherical symmetry in the
appropriate limit. This approach is described elsewhere in axisymmetry
\cite{gentle99b}, and will also be mentioned briefly below.

\begin{figure}[ht]
  \doublefigure{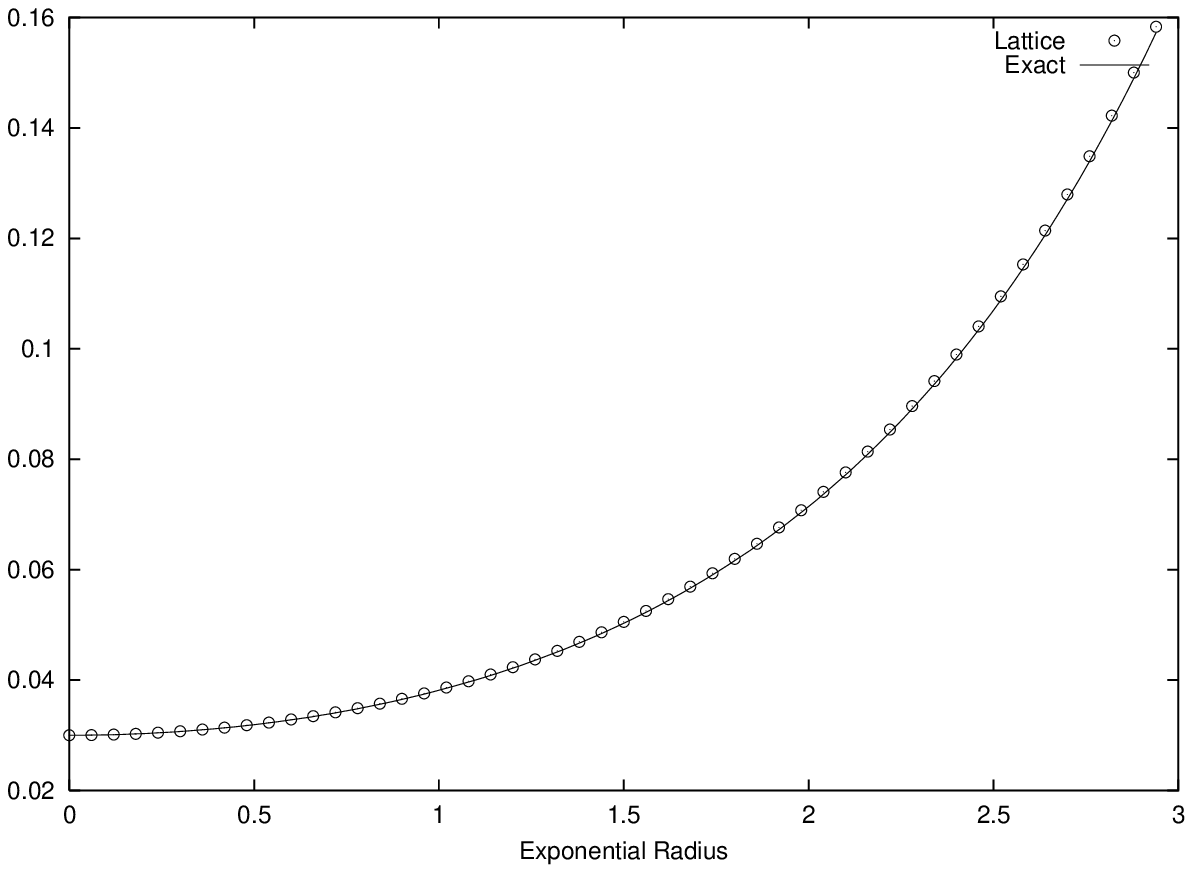}{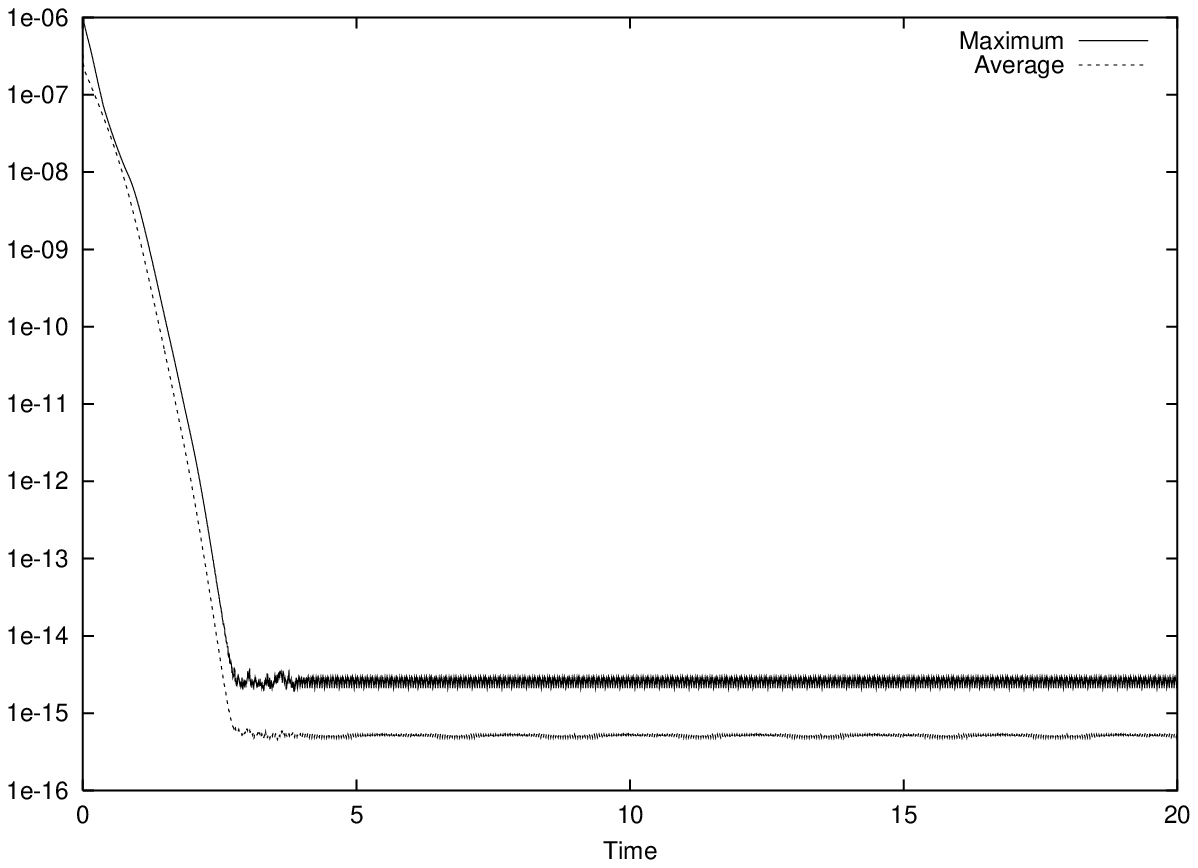}
  \caption{ Evolution of the Schwarzschild solution in
    isotropic coordinates using a lattice of 200 vertices.  {\sl (a)}
    The radial edge length $l$ (circles; every fourth point shown)
    plotted against the exponential radius coordinate $\eta$ at
    $t=10M$.  The exact solution is also shown (solid line). {\sl(b)}
    The time evolution of the fractional rate of change of the radial
    edge lengths.  After an initial period of relaxation, the
    evolution settles down to a static solution.}
  \label{fig:schwarzschild}
\end{figure}

In figure \ref{fig:schwarzschild} we display a sample evolution of
Schwarzschild initial data.  The initial data is expressed in
isotropic coordinates, and the evolution is performed using the
(fixed) analytic lapse and an area locking condition ($\partial _t
g_{\theta\theta}=0$).  The lattice implementation of area locking
demands that the edges which locally span the $\theta$-axis remain
constant, thereby fixing the area of the two-sphere at that vertex.
After an initial relaxation phase, during which the maximum fractional
changes in the edges (metric) are less than $10^{-6}$, the solution
settles into an entirely static configuration.  This application of
lattice gravity demonstrates that the approach can yield accurate,
stable evolutions.

\subsection{Axisymmetric initial data}

\subsubsection{Brill waves}

The axisymmetric vacuum initial value problem first posed by Brill
\cite{brill59} contains gravitational radiation in an otherwise flat
initial three-geometry.  Conformal decomposition using a flat,
simply-connected base metric leads only to trivial solutions; Brill
introduced a metric of the form
\begin{equation}
  \label{eqn:brill_metric}
  dl^2 = \psi^4 \left\{ e^{2q} ( d\rho ^2 +dz^2 ) +
    \rho^2 \, d\phi^2 \right\},
\end{equation}
where the arbitrary function $q(\rho,z)$ can be considered the
distribution of gravitational wave amplitude, and is subject to
certain boundary conditions to ensure that the mass is asymptotically
well define \cite{brill59}. With this choice of background metric, the
Hamiltonian constraint takes the form
\begin{equation}  \label{eqn:brill_hamiltonian}
  \bigtriangledown ^2 \psi = - \frac {\psi}{4} \left( \frac {\partial
      ^2 q}{\partial \rho ^2} + \frac {\partial ^2 q}{\partial z ^2}
  \right)
\end{equation}
which is solved for $\psi (\rho,z)$ once $q(\rho,z)$ is given.

The construction of lattice-based Brill wave initial data was
originally considered by Dubal \cite{dubal89} using a lattice built
from prisms, and later by Gentle \cite{gentle99a,gentle99b} using a
tetrahedral lattice. The approach, modeled on the analysis of Brill,
uses a conformal decomposition technique and solves for the single
conformal factor per vertex.  The lattice is constructed to mirror the
cylindrical polar coordinate system in which the continuum Brill wave
metric is written.  The conformal decomposition is obtained by
integrating spacelike geodesics between vertices of the lattice, and
assigning these lengths directly to the ``base'' lattice edges.

The solutions obtained for the tetrahedral three-geometry were in
excellent agreement with previous finite-difference studies
\cite{gentle99a,gentle99b}, thus confirming both the applicability of
simplicial gravity, and the earlier numerical results.  The
prism-based calculations were considerably less accurate than the
corresponding tetrahedral solution.  Apparent horizons and ADM masses
were also calculated for the tetrahedral initial data, again finding
agreement with previous studies.  The critical wave amplitude at which
an apparent horizon first forms on the initial surface was found to be
in complete agreement with earlier numerical values \cite{gentle99a}.

\subsubsection{Distorted black holes}

In this section we consider another axisymmetric, non-rotating
configuration which contains a moment of time symmetry: the
``distorted black hole'' spacetime first considered by Bernstein
\cite{bernstein93}.  These solutions represent an initial slice
containing a black hole together with Brill waves perturbations, a
natural generalisation of the preceding section, where Brill waves
were considered on a flat Euclidean space.

The distorted black hole spacetime is astrophysically interesting, as
it captures the ``ring-down'' phase of the merger and coalescence of a
binary black hole system.  Following merger, a single distorted
(non-spherical) black hole forms, then evolves towards the static
Schwarzschild solution by the emission of gravitational radiation.
This problem is also amenable to perturbation approaches applied to
the standard black hole solutions, providing a regime in which
numerical relativity, theoretical analysis and gravitational wave
observations combine to shed light on the foundations of general
relativity.
 
The Brill wave plus black hole spacetime \cite{bernstein93} is
obtained by mirroring the original work of Brill \cite{brill59}.  A
perturbation is introduced onto a ``background metric'', and the
conformal factor is calculated from the single initial value equation.
Bernstein wrote the physical metric on the initial surface in the form
\begin{equation}
  \label{eqn:black_hole_brill}
  dl^2 = \psi ^4 \left\{ \mbox{e}^{2q} \left( d\eta ^2 +d\theta ^2\right)  +
    \sin ^2 \theta \, d\phi^2 \right\},
\end{equation}
where the exponential radius coordinate $\eta$ is defined by $\rho = m
\exp (\eta) /2$.  The black hole topology is obtained by demanding
that the two-sphere at $\eta=0$ is a minimal surface (isometry
surface), which connects two asymptotically flat sheets. The ``mass''
$m$ is that of the black hole alone, and corresponds to the ADM mass
measured at spatial infinity when $q(\eta,\theta)=0$.

Distorted back hole initial data has been successfully constructed
using Regge calculus \cite{gentle99b}. It is natural to represent the
three-metric in spherical polar coordinates; the lattice used to model
the distorted black hole spacetime is also matched to the symmetry of
the problem.  Once again, the lattice three-geometry was obtained by
locally aligning the background edges of the lattice with a spherical
polar coordinate system.  This implies, for example, that the $z$-axis
is locally aligned with the azimuthal $\theta$-axis.  This approach
was used as a matter of convenience; there is no requirement that the
lattice be constructed from an underlying coordinate system, but doing
so simplifies both the application of boundary conditions and
comparison with the corresponding solution of the Einstein equations.

The simplicial distorted black hole initial data was found to agree
well with finite-difference calculations, and estimates of the ADM
mass were in excellent agreement with previous studies.  Convergence
estimates indicate that the lattice solution converges as the second
power of the typical lattice discretisation scale towards the true
solution of the Einstein equations \cite{gentle99b}.

\subsubsection{Binary black holes}

\begin{figure}[t]
  \begin{center}
    \begin{tabular}{cc}
      \epsfig{file=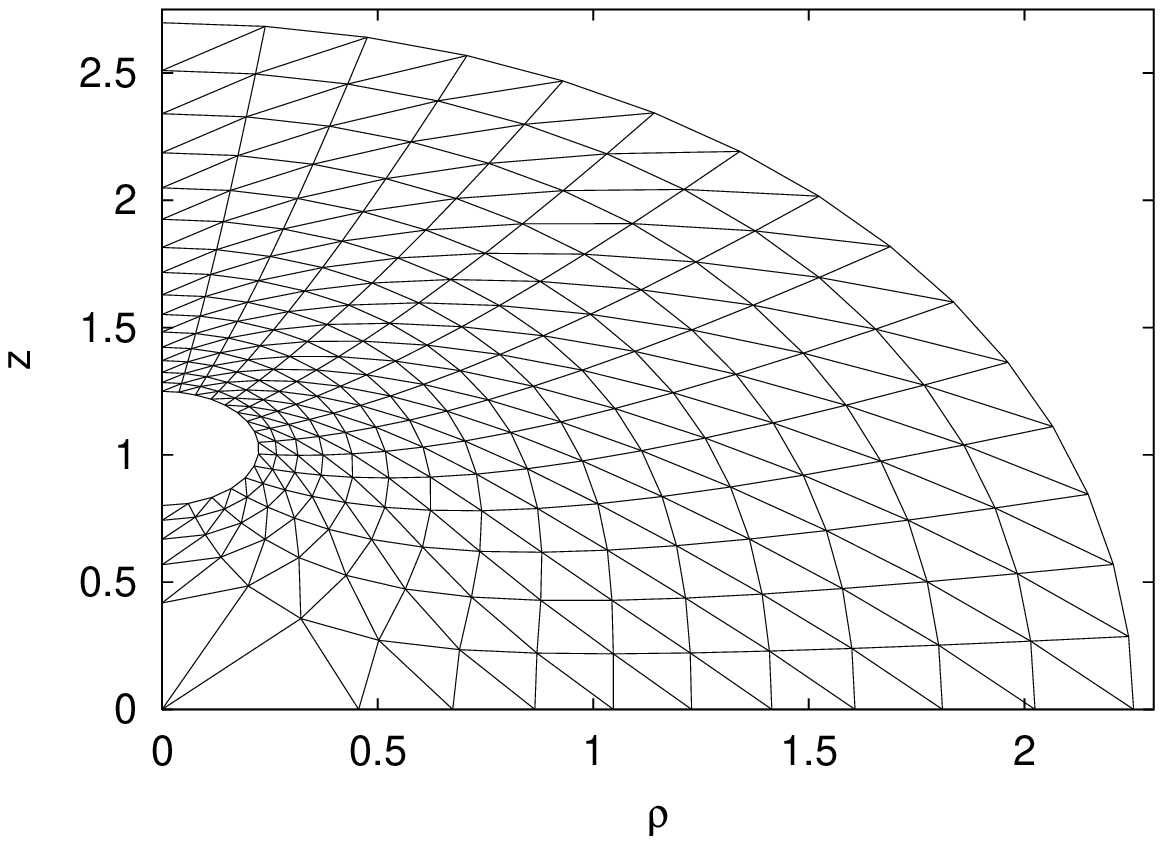, width=0.22\textwidth} &
      \epsfig{file=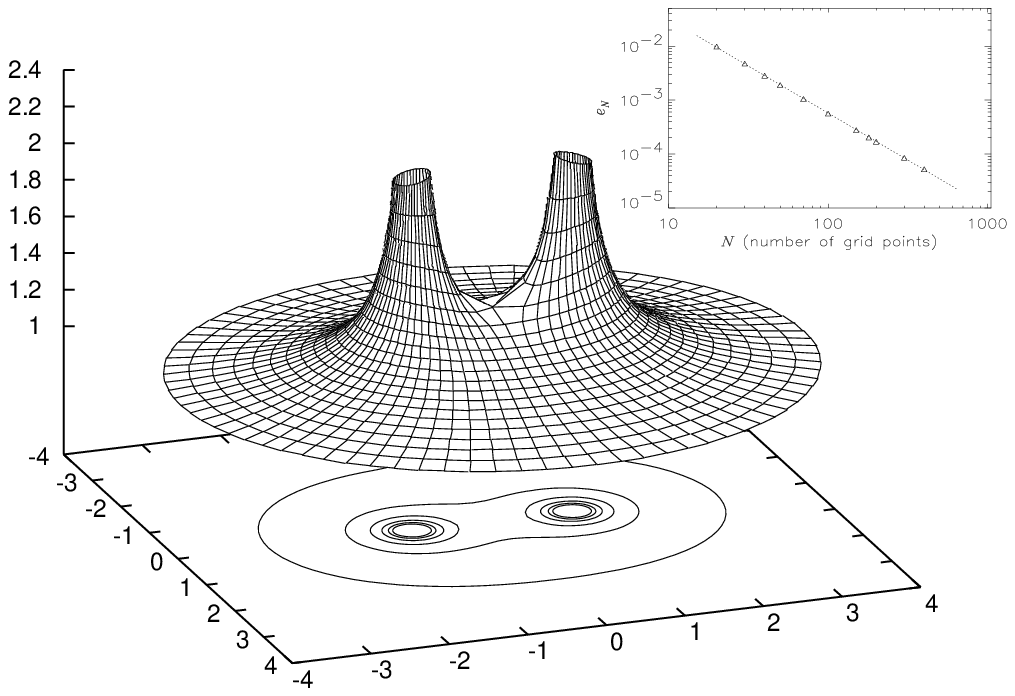, width=0.25\textwidth} \\
      \mbox{\sl(a)} & \mbox{\sl(b)} 
    \end{tabular}
  \end{center}
  \caption{{\sl (a)} A typical example of the lattice used to construct Misner
    initial data for a binary black hole system. The lattice is shown
    close to the $z>0$ throat in $(\rho,z)$ coordinates, {\sl (b)} The
    Regge conformal factor $\psi$ for the Misner binary black hole
    initial data set on the $(\rho,z)$ plane with $\mu = 2.2$.  The
    inner portion of the domain is shown.  The solution is
    obtained on the first quadrant with $501\times 501$ lattice
    vertices and an outer boundary at approximately $r^2 = \rho^2 +
    z^2= 100^2$.  The inset shows the convergence of the simplicial
    solution towards the continuum; the line of best fit has gradient 1.8.}
  \label{fig:misner} 
\end{figure}

Misner \cite{misner60} obtained an axisymmetric solution to the
initial value equations of general relativity suitable for the study
of the head-on collision of equal mass, non-rotating black holes.  Due
to the complexity of the problem there is no known analytic solution
for the time development of this initial data.  

The maximal extension of the standard Schwarzschild black hole
solution involves two asymptotically flat sheets, joined through the
black hole throat.  The spacetime has an isometry through the throat.
The binary black hole solution obtained by Misner \cite{misner60} is
the natural generalisation of this solution; it consists of two
asymptotically flat sheets joined to one another by two throats.
The solution represents two black holes with equal mass and zero
angular momentum at a moment of time-symmetry ($K_{ab}=0$).  The holes
are located on the z-axis at $z=\pm \coth \mu $ and their throats
have radius $a = 1/\sinh\mu$, which also determines the ADM mass of
each individual hole if measured in isolation.  The three-metric takes
the conformally-flat form
\begin{equation}
  \label{eqn:misner}
  dl^2 = \psi^4 \left( d\rho^2 + dz^2 + \rho^2 d\phi^2 \right)
\end{equation}
where the conformal factor is given by
\begin{equation}
  \label{eqn:misner_psi}
  \psi = 1 + \sum_{n=1}^{\infty} \frac{1}{\sinh\left(n\mu\right)}
  \left( \frac{1}{\sqrt{\rho^2 + z_+^2} } + 
    \frac{1}{\sqrt{\rho^2 + z_-^2}} \right)
\end{equation}
and $z_\pm = z \pm z_n$ with $z_n = \coth(n\mu)$.  This is a solution
of the vacuum Hamiltonian constraint at a moment of time symmetry,
$R=0$.  The single free parameter $\mu$ controls the separation, mass
and radius of the black holes, with larger $\mu$ corresponding to
greater separation and smaller bare mass \cite{misner60}.

In this section we present lattice solutions which correspond to the
Misner initial data.  Unlike the axisymmetric initial data presented
in the previous sections, the existence of an analytic solution allows
us to directly evaluate the accuracy and convergence of the lattice
solution.  The lattice is built using \cadez\ coordinates
\cite{smarr76}, which are spherical about both black hole throats and
reduce to standard polar coordinates far from the origin.  This
structure necessarily introduces a coordinate singularity; in \cadez\ 
coordinates the singularity occurs at the origin.

The singular nature of the ``physical'' coordinate system hindered
finite-difference studies of the Misner spacetime \cite{anninos95} ---
we argue that the lattice approach overcomes this problem.  While we
are free to use the \cadez\ coordinates to construct the initial
lattice (``location of vertices''; see figure \ref{fig:misner}(a)),
the resulting lattice consists entirely of scalar edge lengths.
Vertex positions are assigned (for example) using a regularly spaced
grid in the \cadez\ coordinate system, and then integration of
geodesics is used to assign the background edge lengths of the lattice
given the flat background metric.  From this point on the simplicial
simulation consists entirely of evaluating scalar functions of the
edge lengths.  There is no singularity in the lattice construction.

A typical Regge solution for the conformal factor $\psi$ is shown in
figure \ref{fig:misner}(b), with $\mu =2.2$.  This parameter is used
to specify the centre ($z=\pm \cosh \mu$) and radius ($a=\mbox{csch}\ 
\mu$) of the black holes. This choice guarantees that the initial data
under consideration is of the Misner type, although it is clearly
possible to investigate other configurations once the code has been
benchmarked on the analytic solution.  

The lattice solution shown in figure \ref{fig:misner}(b) is found to
be in good agreement with the analytic solution, and shows almost
second order convergence towards the analytic solution. The inset to
figure \ref{fig:misner}(b) shows that the convergence rate is
approximately 1.8.  We find that as the number of vertices $N$ along
each axis is increased, lattice edges far from the holes scale as
$1/N$.  Near the saddle point, however, the edges scale as
$1/\sqrt{N}$.  It is precisely this behaviour which leads to the
(slightly) anomalous convergence rate; the slowly converging terms
near the origin dominate.  This ``problem'', inherited from the
singularity in the \cadez\ coordinates, can be easily overcome by
further refinement of the lattice near the saddle point.

Despite these minor issues, the Regge approach recovers a second-order
accurate approximation to the solution of Einstein's equations for the
Misner binary black hole initial data.  Following further refinement
of the lattice near the saddle point, the method should yield precisely
second order convergence.  The \cadez\ coordinates are used only as a
convenience when constructing the initial lattice.  If and when such
an approach proves problematic, the lattice can, and should, be
further subdivided to obtain a consistent ``triangulation''.

\subsection{Brill wave evolutions}

\begin{figure}[t]
  \doublefigure{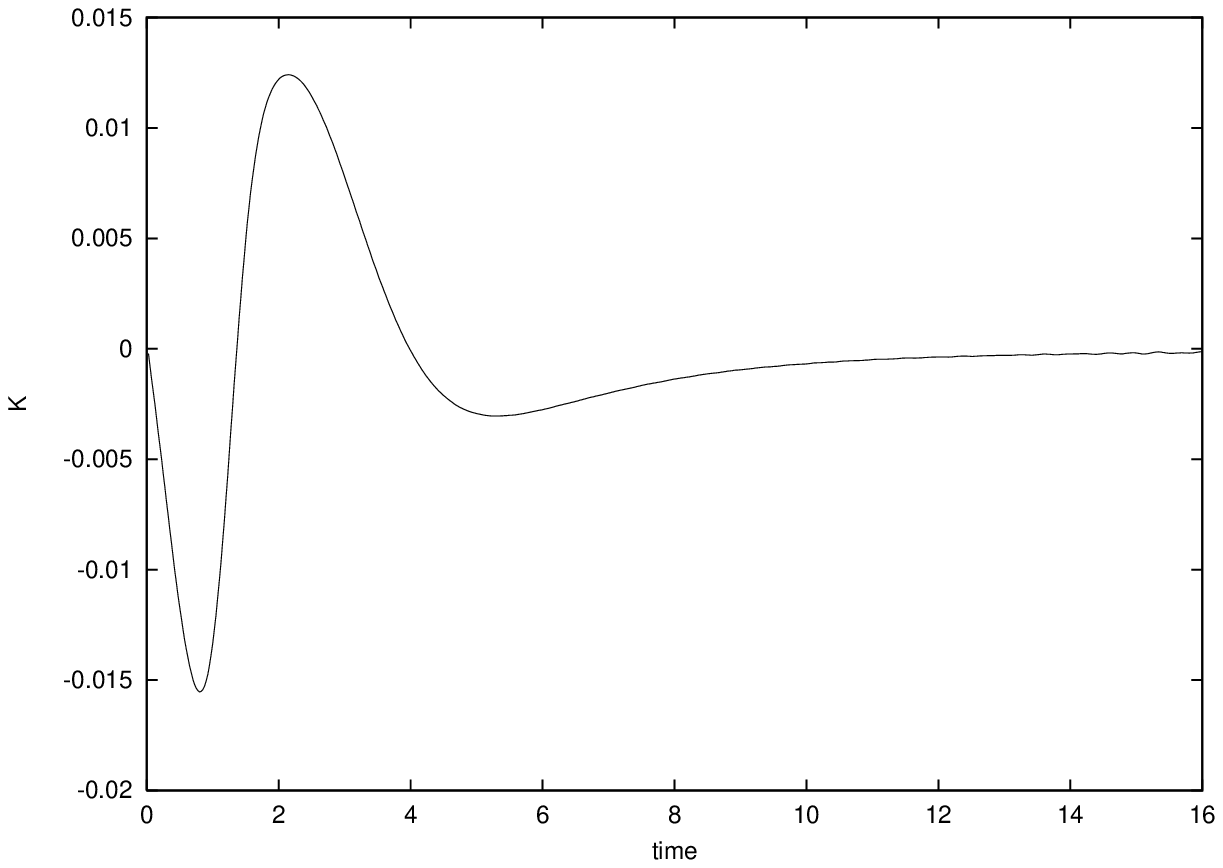}{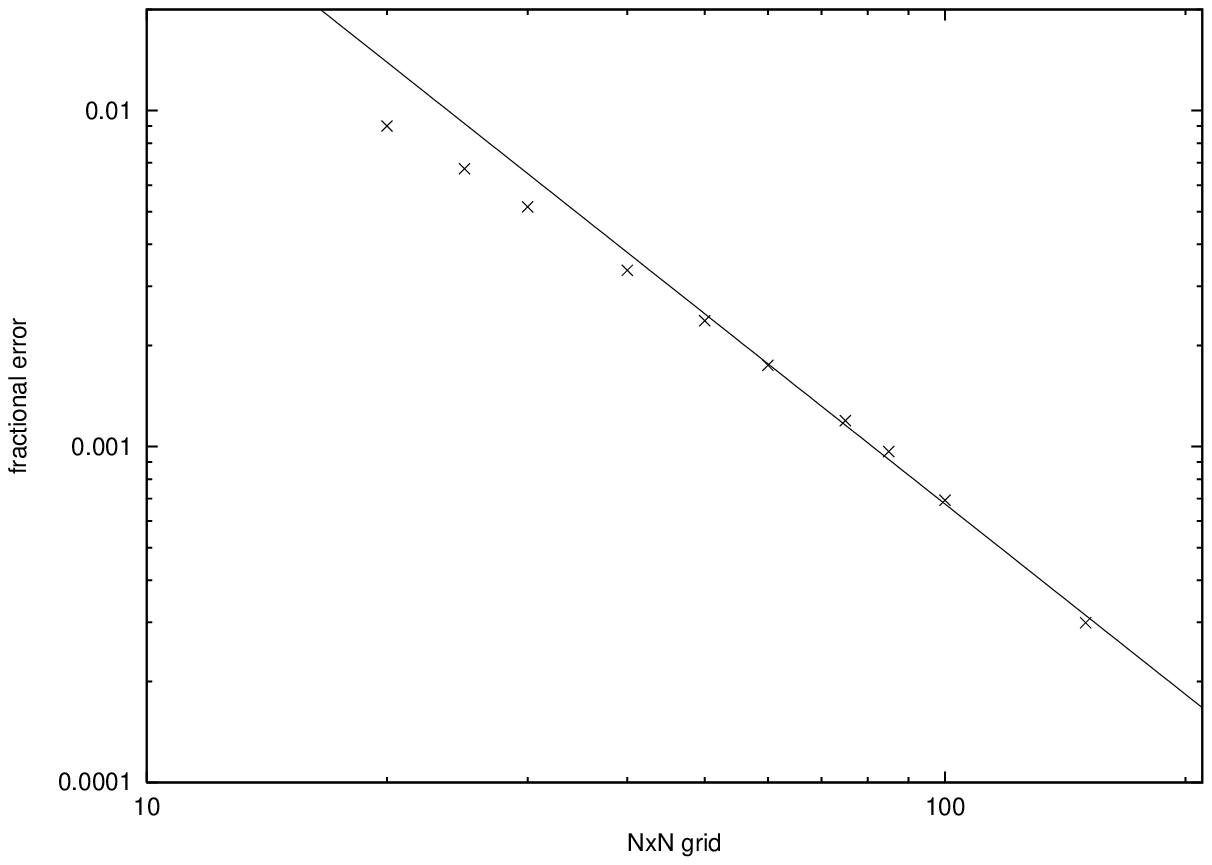}
  \caption{Brill wave evolution. {\sl (a)} Time evolution of $K =
    K^{a}_{\ a}$, evaluated at the origin.  We use the fact that $K$
    is equivalent to the fractional rate of decrease of the volume per
    vertex.  Evolution is shown for a wave of amplitude $a=0.01$ on a
    $100\times 100$ grid, timestep $\Delta t = \Delta \rho/4$, and
    with the outer boundary at $\rho=z=10$.  {\sl(b)} Mean fractional
    difference between magnitude of the first ``bump'' in {\sl(a)} and
    the corresponding ``continuum value'', plotted as a function of
    the grid resolution $N\times N$. The continuum value is estimated
    using Richardson extrapolation of the two most accurate values.
    We observe close to second order convergence; the gradient of the
    line of best fit is approximately $1.9$.}
    \label{fig:evolution}
\end{figure}

In this section we describe an application of lattice gravity to the
evolution of the (low amplitude) Brill wave initial data discussed
above.  Although all calculations are performed using lattice gravity,
in this section we describe the techniques and gauge choices in the
standard language of numerical relativity.  Although the Regge
framework is complete and self-contained, we find it useful to
highlight connections with the continuum and more ``standard''
approaches to numerical relativity.

The spatial metric is written in cylindrical polar coordinates, and
assumed to take the form
\begin{equation}
  \label{eqn:metric}
  ds^2 = \psi^4 \left\{ \, \xi^2 \left( d\rho^2 + dz^2 \right) 
        + \rho^2 d\phi^2 \, \right\}
\end{equation}
throughout the evolution, where $\xi = \xi(\rho,z,t)$ and $\psi =
\psi(\rho,z,t)$.  The lattice is adapted to these coordinates, as
described elsewhere \cite{gentle99a}. This form of the three-metric
$\gamma_{ab}$ was also chosen by Garfinkle and Duncan
\cite{garfinkle01}, and implicitly involves the imposition of the two
conditions
\begin{equation}
  \label{eqn:gauge}
        \gamma_{\rho\rho} = \gamma_{zz}, \qquad \mbox{and} \qquad
\gamma_{\rho z} = 0,
\end{equation}
in addition to the initial assumption of axisymmetry.  These two
conditions are effectively a choice of gauge, and through the Einstein
equations, determine the shift vector $\beta^i$ \cite{garfinkle01}.
Whilst the same holds true in the lattice approach, we follow a
different strategy in this initial evolution of gravitational
radiation on a lattice.

The conditions (\ref{eqn:gauge}) are applied directly to the lattice
edges corresponding to the spatial components of the three-metric.  A
zero shift condition, $\beta^i=0$, is also implemented on the lattice
to ensure that the timelike worldlines generated by the vertices are
orthogonal to the lower hypersurface.  This is an over-specification
of the available degrees of freedom, but we argue that for weak Brill
waves ($a<<1$) the spacetime is a perturbation of flat space, and thus
all of the above conditions can be enforced to within a reasonable
accuracy.  The choice of zero shift, together with the extra gauge
choice (\ref{eqn:gauge}), will in general violate the Einstein
equations and their simplicial counterparts.  For small perturbations
away from flat space we expect that these errors are controllable. The
generic evolution of medium to large amplitude waves will require a
consistent specification of the available gauge freedoms.  The choices
outlined above, however, are sufficient to perform the initial
low-amplitude test evolutions presented in this paper.

The remaining gauge freedom is determined by setting $\alpha=1$ (unit
lapse).  This condition is not ideal, as it is known that even weak
waves will impart non-zero velocities to the vertices of the lattice
(or grid).  Once the wave has propagated away from the centre of the
grid we expect to recover a non-trivial, non-static coordinatisation
of flat space. For strong amplitude waves we will require more
advanced slicing conditions, such as maximal ($\mbox{Tr}K=0$) or
harmonic slicing.  Geodesic slicing is sufficient, however, for this
initial low-amplitude test problem.

Boundary conditions are imposed on the lattice to enforce the
reflection symmetry about the $\rho=0$ axis, together with reflection
symmetry about the $z=0$ axis to simplify the computations.  These
conditions are implemented using ghost cells centred about the axis.
This approach has been used successfully in previous studies
\cite{garfinkle01}, and is found to work well.  The outer boundary
conditions are implemented using a radiative ``Sommerfeld'' condition,
implemented in differential form on $\psi$ and $\xi$ along the outer
boundary  \cite{alcubierre00}.

\begin{figure}
  \quadfigure{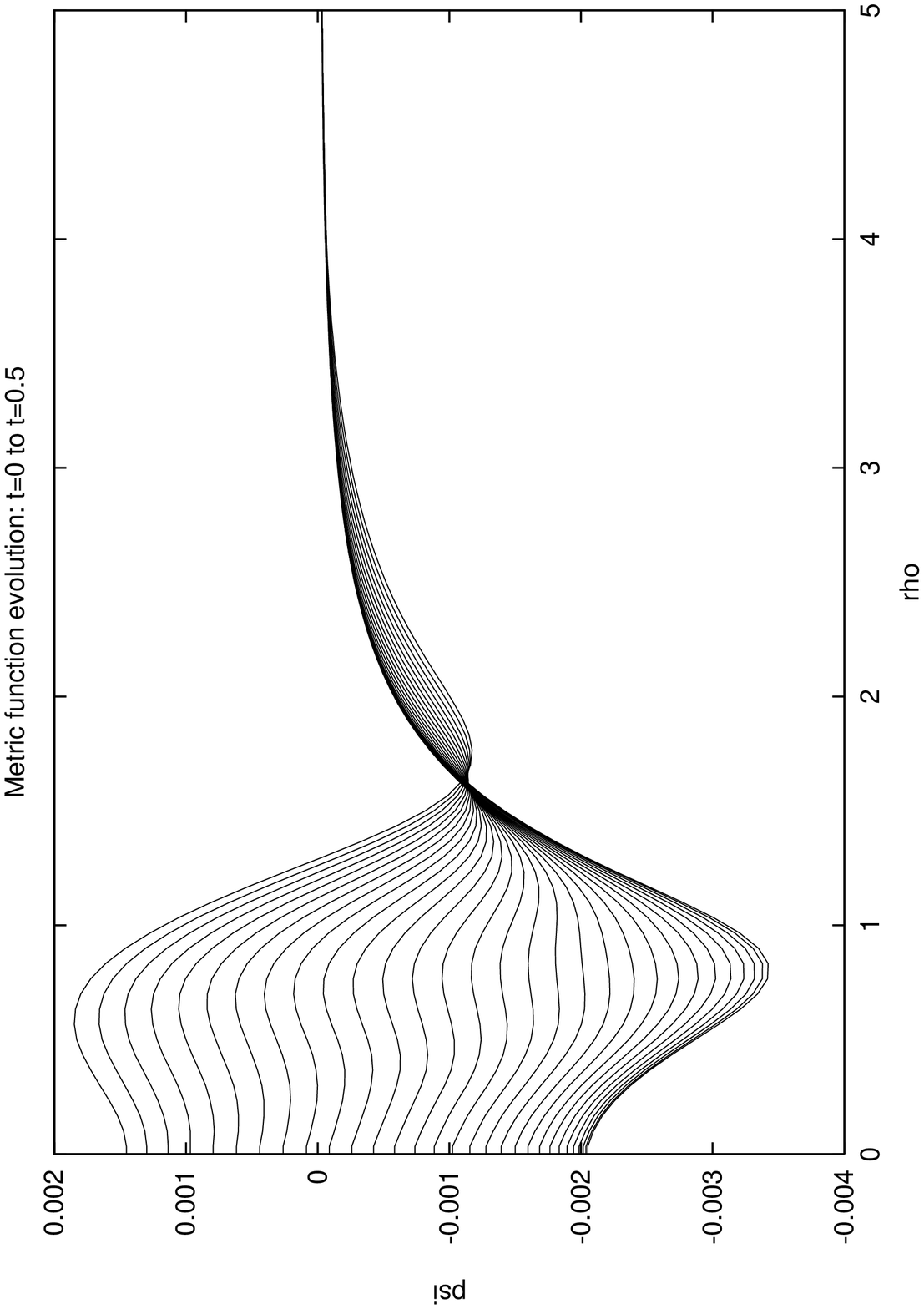}{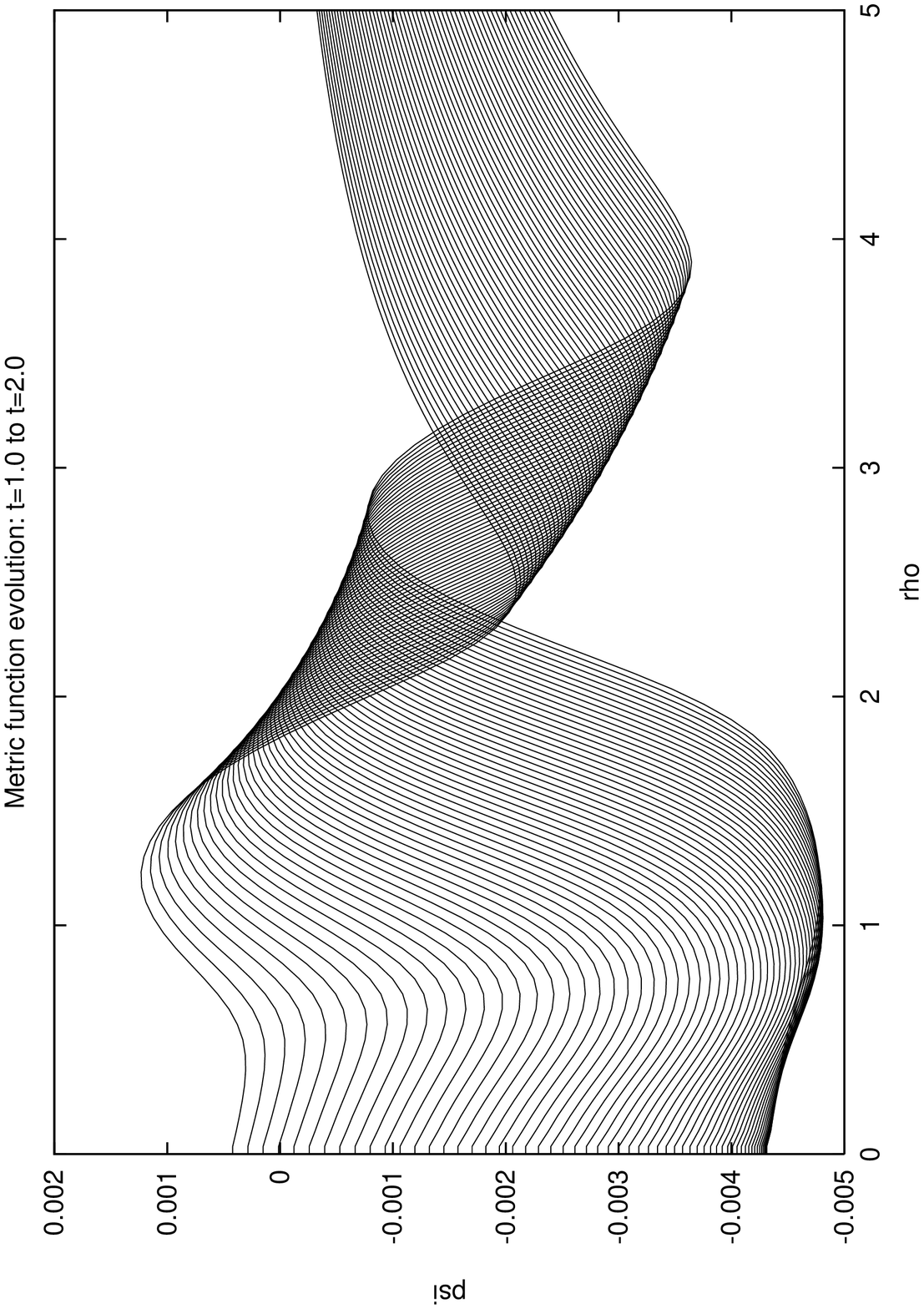}{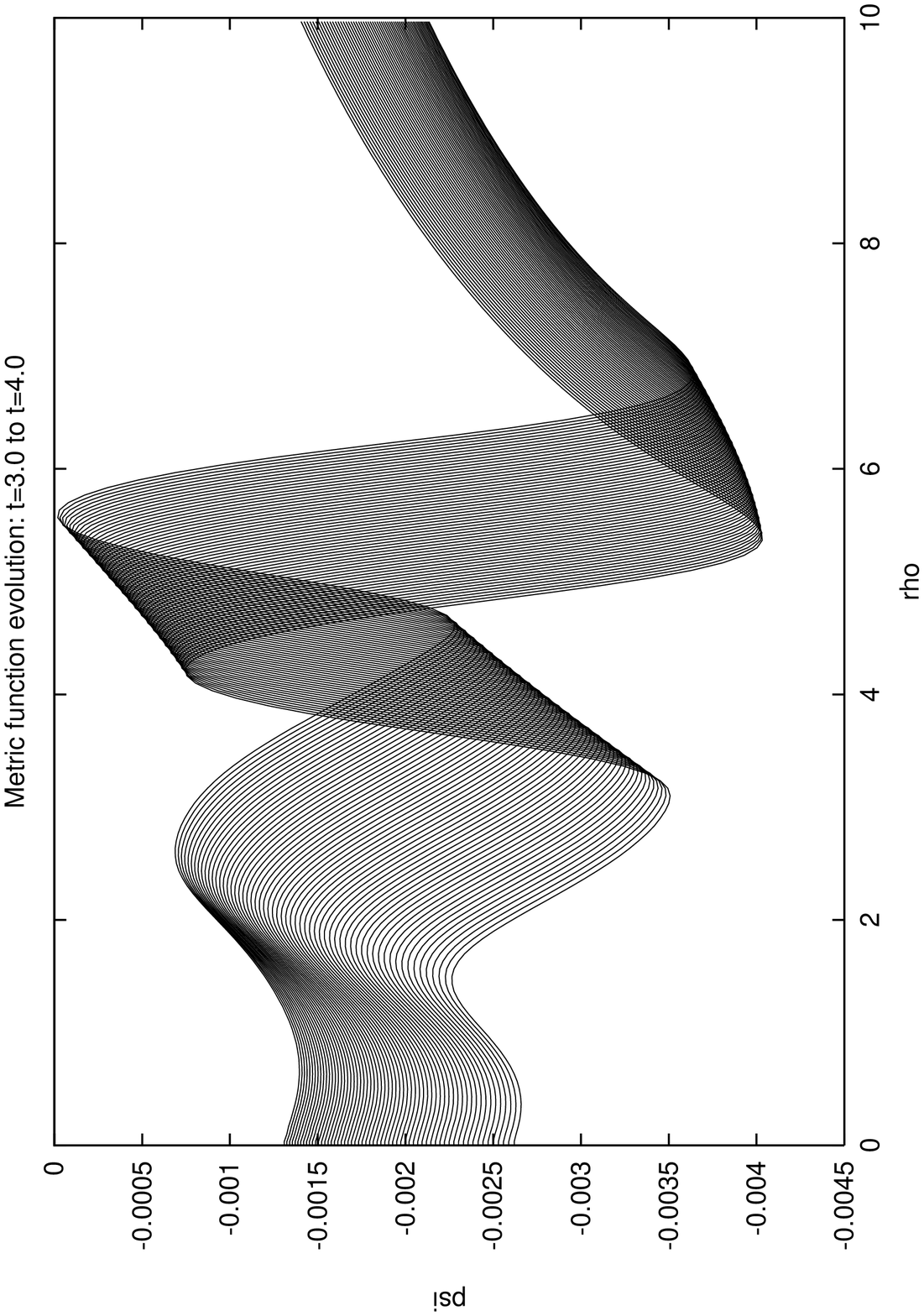}{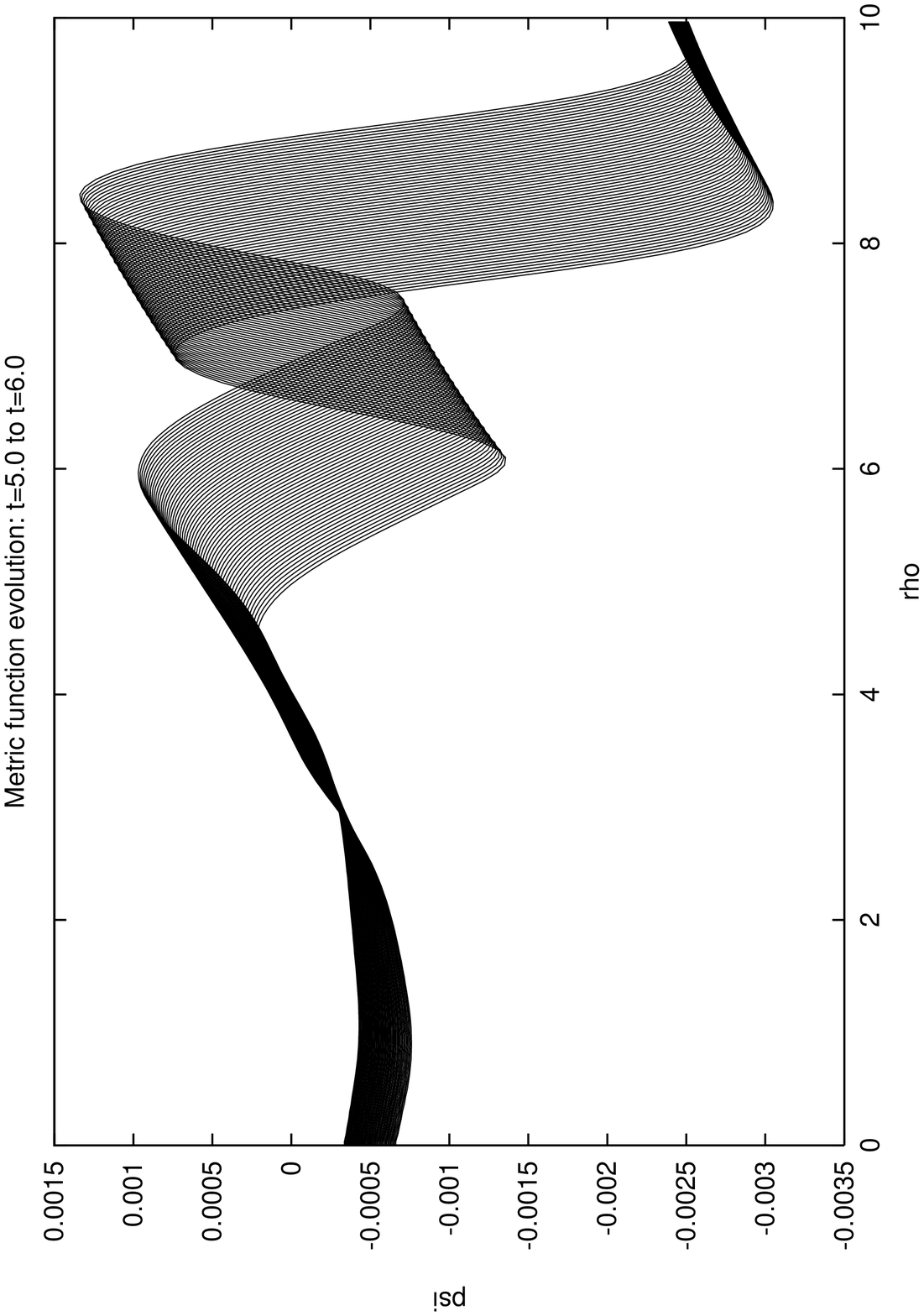}
  \caption{The deviation of the conformal factor, $\psi^2-1$,
    plotted as a function of $\rho$ along the line $z=0$ for various
    times ({\sl (a)} $t=0$ to $0.5$, {\sl (b)} $t=1$ to $2$, {\sl (c)}
    $t=3$ to $4$.  {\sl (d)} $t=5$ to $6$) for a Brill wave with
    amplitude $a=0.01$.  We use the ``Eppley-type'' Brill function
    $q(\rho,z)$ \cite{eppley77}, and show solutions for a lattice with
    $100\times 100$ vertices. Note that the x-axis scale changes in
    parts {\sl (c)} and {\sl(d)}, and the outer boundary is placed at
    $\rho=z=10$.} \label{fig:psi}
\end{figure}

Results are shown in figures \ref{fig:evolution} and \ref{fig:psi},
and demonstrate that Regge calculus can successfully and accurately
propagate a gravitational wave through the lattice.  An initial
convergence estimate is performed by examining the convergence of the
magnitude of the first ``bump'' in figure \ref{fig:evolution}(a).
This feature of the solution is found to converge at close to second
order, demonstrating a degree of consistency in the solutions
obtained.  Figure \ref{fig:psi} displays the deviation of the
conformal factor ($\psi^2-1$) at various times; the propagation of the
wave can be observed, with the inner region returning to a roughly
constant value (``flatness'') once the wave has escaped towards the
boundary.  Reflections from the outer boundary currently prevent long
term evolutions.

\section{The future of lattice approaches to numerical relativity}

We have reviewed several recent applications of Regge calculus, and
demonstrated that it provides a unique, and thus far successful, alternative
to the more standard techniques employed in numerical relativity.

In particular, we have presented new solutions of lattice gravity
corresponding to initial data for the head-on collision of black
holes, together with the first successful time evolution of
gravitational radiation on a lattice.  Despite these and the other
studies described above, significant questions remain regarding the
fundamental structure of Regge calculus.

Much work remains to be done to address the dual and diffeomorphic
structures of the lattice, the relation of this and other lattice
approaches to standard finite element and finite volume
discretisations of differential equations, and the inclusion of
matter. On the numerical front, further applications in three-plus-one
dimensions are vital, requiring the development of a modern parallel
code.

Despite these issues, every indication is that the lattice approached
to gravity developed by T.~Regge will continue to provide an
interesting and complementary approach to numerical relativity.

\section*{Acknowledgments}
The author is grateful to the organisers of the Third Australasian
Conference on General Relativity and Gravitation (held in Perth,
Western Australia July 11-13 2001) for the invitation to present the
talk on which this material is based.  The author is also grateful for
the endless enthusiasm and encouragement provided by Warner Miller and
Leo Brewin during his work on lattice gravity.  Arkady Kheyfets and
Matthew Anderson also provided helpful comments and suggestions during
the completion of parts of this work.

\gdef\journal#1, #2, #3, #4 {{#1}, {\bf #2}, #3 {(#4)}. }
\gdef\PR{\it{Phys. Rev.}}
\gdef\NC{\it{Nouvo~Cimento}}
\gdef\CQG{\it{Class. Quant. Grav.}}
\gdef\JMP{\it{J. Math. Phys.}}
\gdef\PTP{\it{Prog. Theor. Phys.}}
\gdef\AP{\it{Ann. Phys.}}
\gdef\IJTP{\it{Int. J. Theor. Phys.}}

\end{document}